\documentstyle[12pt]{article}                    %%%
\newfont{\ffont}{msym10}                        %%
\newcommand{\beq}{\begin{equation}}             %%
\newcommand{\eeq}{\end{equation}}               %%
\newcommand{\bqry}{\begin{eqnarray}}            %%
\newcommand{\eqry}{\end{eqnarray}}              %%
\newcommand{\bqryn}{\begin{eqnarray*}}          %%
\newcommand{\eqryn}{\end{eqnarray*}}            %%
\newcommand{\NL}{\nonumber \\}                  %%
\newcommand{\preprint}[1]{\begin{table}[t]      %%
            \begin{flushright}                  %%
            \begin{large}{#1}\end{large}        %%
            \end{flushright}                    %%
            \end{table}}                        %%
\newcommand{\PD}[2]                             %%
    {\frac{\partial^{#2}}{\partial #1^{#2}}}    %%
\begin{document}
\preprint{LA-UR-97-2333 \\ TAUP-XXXX-97}
\title{New Quadratic Baryon Mass Relations}
\author{\\ L. Burakovsky\thanks{E-mail: BURAKOV@PION.LANL.GOV}, \  
T. Goldman\thanks{E-mail: GOLDMAN@T5.LANL.GOV} \
\\  \\  Theoretical Division, MS B285 \\  Los Alamos National Laboratory \\ 
Los Alamos, NM 87545, USA \\  \\  and  \\  \\
L.P. Horwitz\thanks{E-mail: HORWITZ@TAUNIVM.TAU.AC.IL. Also at Department of 
Physics, Bar-Ilan University, Ramat-Gan, Israel } \
\\  \\ School of Physics and Astronomy \\ Tel-Aviv University \\ Ramat-Aviv,
69978 Israel \\}
\date{ }
\maketitle
\begin{abstract}
By assuming the existence of (quasi)-linear baryon Regge trajectories, we 
derive new quadratic Gell-Mann--Okubo type baryon mass relations. These 
relations are used to predict the masses of the charmed baryons absent from 
the Baryon Summary Table so far, in good agreement with the predictions of 
many other approaches.
\end{abstract}

\bigskip

\section*{  }
The investigation of the properties of hadrons containing heavy quarks is of
great interest for understanding the dynamics of the quark-gluon interaction.
Recently predictions about the heavy baryon mass spectrum have become a 
subject of increasing interest $[1-11],$ due to current experimental 
activity of several groups at CERN \cite{WA89}, Fermilab \cite{E687} and 
CESR \cite{CLEO,CLEO1} aimed at the discovery of the baryons so far absent 
from the Baryon Summary Table \cite{data}. Recently, for the LHC, B-factories 
and the Tevatron with high luminosity, several experiments have been proposed 
in which a detailed study of heavy baryons can be performed. In this 
connection, an accurate theoretical prediction for the baryon mass spectrum 
becomes a guide for experimentalists. To calculate the heavy baryon mass 
spectrum, potential models $[1,\;17-22],$ nonrelativistic quark models 
$[23-25],$ relativistic quark models \cite{Ebert}, bag models $[26-29],$ 
lattice QCD $[30-32],$ QCD spectral sum rules \cite{QSSR}, heavy quark 
effective theory $[11,\;34-36],$ chiral perturbation theory \cite{Sav}, chiral
quark model \cite{GR}, $SU(4)$ skyrmion model \cite{RRS}, group theoretical
\cite{Leb,Chan,SA} and other approaches $[3-5,\;7,\;8,\;40-44]$ are widely 
used.   

The charm baryon masses measured to date are\footnote{For $\Sigma ^\ast _c$
we take the uncertainty weighted average of the results of ref. \cite{Amm}, 
$2530\pm 5\pm 5$ MeV, and the most recent results by CLEO \cite{CLEO1}, 
$2518.6\pm 2.2$ MeV.} 
\cite{data}
\bqryn
\Lambda _c & = & 2285\;{\rm MeV,} \\
\Sigma _c & = & 2453\pm 1\;{\rm MeV,} \\
\Xi _c & = & 2468\pm 2\;{\rm MeV,} \\
\Omega _c & = & 2704\pm 4\;{\rm MeV,} \\
\Sigma ^\ast _c & = & 2521\pm 4\;{\rm MeV,} \\
\Xi ^\ast _c & = & 2644\pm 2\;{\rm MeV.}
\eqryn
An observation of the $\Xi ^{'}_c=2563\pm 15$ MeV was reported by the WA89
Collaboration \cite{WA89}. The $\Omega ^\ast _c,$ as well as double- and
triple-charmed baryons, have not yet been observed. 

Almost all very recent calculations very consistently predict the mass of the 
$\Xi ^{'}_c$ to be around 2580 MeV \cite{Sav,RLP,LRP,ZZ,Fran,Jen} (see also 
\cite{Itoh,DDV,ST}). Similarly, the mass of the $\Omega ^\ast _c$ is very
consistently predicted to be around 2770 MeV \cite{Sav,Ros,RLP,LRP,ZZ,GR,Jen} 
(see also \cite{MR,FR,Ponce,SM,DDV}). Predictions for the double- and 
triple-charmed baryon masses are less definite.  

Here we wish to extend the approach based on the assumption of 
(quasi)-linearity of the Regge trajectories of heavy hadrons in the low-energy
region, initiated in our previous papers for heavy mesons \cite{BGH1,BGH2}, to 
baryons. We shall show that new quadratic Gell-Mann--Okubo type baryon mass 
relations can be obtained, and used to predict the missing charmed baryon 
masses. As we shall see, the predicted masses are in good agreement with the 
results of many other approaches, which should add confidence to an 
experimental focus on the predicted ranges.

Let us assume, as in \cite{BGH1,BGH2}, the (quasi)-linear form of Regge 
trajectories for baryons with identical $J^P$ quantum numbers (i.e., belonging 
to a common multiplet). Then for the states with orbital momentum $\ell $ one 
has $(i,j,k$ stand for the corresponding flavor content)
\bqryn
\ell & = & \alpha ^{'}_{kii}m^2_{kii}\;+a_{kii}(0), \\    
\ell & = & \alpha ^{'}_{kji}m^2_{kji}\;\!+a_{kji}(0), \\    
\ell & = & \alpha ^{'}_{kjj}m^2_{kjj}+a_{kjj}(0).    
\eqryn
Using now the relation among the intercepts \cite{inter,DB,Kai},
\beq
a_{kii}(0)+a_{kjj}(0)=2a_{kji}(0),
\eeq
one obtains from the above relations
\beq
\alpha ^{'}_{kii}m^2_{kii}+\alpha ^{'}_{kjj}m^2_{kjj}=
2\alpha ^{'}_{kji}m^2_{kji}.
\eeq
In order to eliminate the Regge slopes from this formula, we need a relation 
among the slopes. Two such relations exist,
\beq
\alpha ^{'}_{kii}\cdot \alpha ^{'}_{kjj}=\left( \alpha ^{'}_{kji}\right) ^2,
\eeq
which follows from the factorization of residues of the $t$-channel poles
\cite{Pas,Igi,KY}, and
\beq
\frac{1}{\alpha ^{'}_{kii}}+\frac{1}{\alpha ^{'}_{kjj}}=
\frac{2}{\alpha ^{'}_{kji}},
\eeq
which may be derived by generalizing the corresponding relation for quarkonia 
based on topological expansion and the $q\bar{q}$-string picture \cite{Kai} to
the case of a baryon viewed as a quark-diquark-string object\footnote{This 
structure is known to be responsible for the slopes of baryon trajectories 
being equal to those of meson trajectories \cite{Egu,Sim,RS}.} \cite{prep}.  

For light baryons (and small differences in the $\alpha ^{'}$ values), there
is no essential difference between these two relations; viz., for $\alpha ^{'
}_{kji}=\alpha ^{'}_{kii}/(1+x),$ $x\ll 1,$ Eq. (4) gives $\alpha ^{'}_{kjj}=
\alpha ^{'}_{kii}/(1+2x),$ whereas Eq. (3) gives $\alpha ^{'}_{kjj}=\alpha ^{
'}_{kii}/(1+x)^2\approx \alpha ^{'}/(1+2x),$ i.e, essentially the same result 
to order $x^2.$ However, for heavy baryons (and expected large differences 
from the $\alpha ^{'}$ values for the light baryons) these relations are 
incompatible; e.g., for $\alpha ^{'}_{kji}=\alpha ^{'}_{kii}/2,$ Eq. (3) will 
give $\alpha ^{'}_{kjj}=\alpha ^{'}_{kii}/4,$ whereas from Eq. (4), $\alpha ^{
'}_{kjj}=\alpha ^{'}_{kii}/3.$ One therefore has to choose between these 
relations in order to proceed further. Here, as in \cite{BGH1,BGH2}, we use 
Eq. (4), since it is much more consistent with (2) than is Eq. (3), which we 
tested by using measured light-quark baryon masses in Eq. (2). Kosenko and 
Tutik \cite{KT} used the relation (3) and obtained much higher values for the 
charmed baryon masses than the measured ones (e.g., $\Omega _c=2788$ MeV) and 
those predicted by most other approaches (see Table I). The reason for this is
that lower values for the Regge slopes, as illustrated by the example above, 
lead to higher values for the masses. We shall justify our choice of Eq. (4) 
in more detail in a separate publication \cite{prep}.

It is easy to see that the following relation solves Eq. (4):\footnote{The
notation has changed here, as compared to Eqs. (1)-(4); e.g., $a_{nnn}(0)\equiv
a_{3,0,0}(0),$ $a_{snn}(0)\equiv a_{2,1,0}(0),$ etc.}
\bqry
a^\star _{i_n,j_s,k_c}(0) & = & a^\star (0)-\lambda ^\star _sj_s-\lambda ^\star
_ck_c,\;\;\;a^\star (0)\equiv a^\star _{3,0,0}(0), \\
\frac{1}{\alpha ^{'}_{\star ;i_n,j_s,k_c}} & = & \frac{1}{\alpha _\star ^{'}}+
\gamma ^\star _sj_s+\gamma ^\star _ck_c,\;\;\;\alpha _\star ^{'}\equiv \alpha 
^{'}_{\star ;3,0,0},\;\;\;i_n+j_s+k_c=3,
\eqry
where $i_n,j_s,k_c=1,2,3$ are the numbers of $n$-, $s$-, and $c$-quarks, 
respectively, which constitute the baryon, and the sub- and superscript
$\star $ allows for possible differences between multiplets (such as $\frac{1
}{2}^{+}$ octet and $\frac{3}{2}^{+}$ decuplet).

It then follows from (6) that
\bqry
\alpha ^{'}_\Lambda & = & \alpha ^{'}_\Sigma \;\;=\;\;\frac{\alpha ^{'}_N}{1+
\gamma _s^N\alpha ^{'}_N}, \\  
 &   & \alpha ^{'}_\Xi \;\;=\;\;\frac{\alpha ^{'}_N}{1+2\gamma _s^N\alpha ^{'}_
N}, \\
\alpha ^{'}_{\Lambda _c} & = & \alpha ^{'}_{\Sigma _c}\;=\;\;\!\frac{\alpha ^{'
}_N}{1+\gamma_c^N\alpha ^{'}_N}, \\  
\alpha ^{'}_{\Xi _c} & = & \alpha ^{'}_{\Xi ^{'}_c}\;=\;\;\!\frac{\alpha ^{'}_
N}{1+(\gamma _s^N+\gamma _c^N)\alpha ^{'}_N}, \\
 &   & \alpha ^{'}_{\Omega _c}\;=\;\;\frac{\alpha ^{'}_N}{1+(2\gamma _s^N+
\gamma _c^N)\alpha ^{'}_N}, \\  
 &   & \alpha ^{'}_{\Xi _{cc}}=\;\;\frac{\alpha ^{'}_N}{1+2\gamma _c^N\alpha 
^{'}_N}, \\
 &   & \alpha ^{'}_{\Omega _{cc}}=\;\;\frac{\alpha ^{'}_N}{1+(\gamma _s^N+2
\gamma _c^N)\alpha ^{'}_N},
\eqry
where we use $\star =N$ to represent the $\frac{1}{2}^{+}$ multiplet, and with
$\star =\Delta $ to represent the $\frac{3}{2}^{+}$ multiplet,
\bqry
\alpha ^{'}_{\Sigma ^\ast } & = & \frac{\alpha ^{'}_\Delta }{1+\gamma _s^\Delta
\alpha ^{'}_\Delta }, \\  
\alpha ^{'}_{\Xi ^\ast } & = & \frac{\alpha ^{'}_\Delta }{1+2\gamma _s^\Delta 
\alpha ^{'}_\Delta }, \\
\alpha ^{'}_\Omega & = & \frac{\alpha ^{'}_\Delta }{1+3\gamma _s^\Delta
\alpha ^{'}_\Delta }, \\  
\alpha ^{'}_{\Sigma _c^\ast } & = & \frac{\alpha ^{'}_\Delta }{1+\gamma_c^
\Delta \alpha ^{'}_\Delta }, \\  
\alpha ^{'}_{\Xi ^\ast _c} & = & \frac{\alpha ^{'}_\Delta }{1+(\gamma _s^\Delta
+\gamma _c^\Delta )\alpha ^{'}_\Delta }, \\
\alpha ^{'}_{\Omega _c^\ast } & = & \frac{\alpha ^{'}_\Delta }{1+(2\gamma _s^
\Delta +\gamma _c^\Delta )\alpha ^{'}_\Delta }, \\  
\alpha ^{'}_{\Xi ^\ast _{cc}} & = & \frac{\alpha ^{'}_\Delta }{1+2\gamma _c^
\Delta \alpha ^{'}_\Delta }, \\
\alpha ^{'}_{\Omega _{cc}^\ast } & = & \frac{\alpha ^{'}_\Delta }{1+(\gamma _s^
\Delta +2\gamma _c^\Delta )\alpha ^{'}_\Delta }, \\
\alpha ^{'}_{\Omega _{ccc}} & = & \frac{\alpha ^{'}_\Delta }{1+3\gamma _c^
\Delta \alpha ^{'}_\Delta }.
\eqry

Consider first the $J^P=\frac{3}{2}^{+}$ baryons. Introduce, for simplicity,
\beq
x\equiv \gamma _s^\Delta \alpha ^{'}_\Delta,\;\;\;y\equiv \gamma _c^\Delta 
\alpha ^{'}_\Delta .
\eeq
It then follows from (5)-(13) that 
\bqry
\Delta ^2 & = & \frac{\Sigma ^{\ast 2}}{1+x}-\lambda _s^\Delta \;=\;\frac{\Xi 
^{\ast 2}}{1+2x}-2\lambda _s^\Delta \;=\;\frac{\Omega ^2}{1+3x}-3\lambda _s^
\Delta  \NL
 & = & \frac{\Sigma _c^{\ast 2}}{1+y}-\lambda _c^\Delta \;=\;\frac{\Xi _c^{
\ast 2}}{1+x+y}-\lambda _s^\Delta -\lambda _c^\Delta \;=\;\frac{\Omega _c^{
\ast 2}}{1+2x+y}-2\lambda _s^\Delta -\lambda _c^\Delta  \NL
 & = & \frac{\Xi _{cc}^{\ast 2}}{1+2y}-2\lambda _c^\Delta \;=\;\frac{\Omega _{
cc}^{\ast 2}}{1+x+2y}-\lambda _s^\Delta -2\lambda _c^\Delta  \NL
 & = & \frac{\Omega _{ccc}^2}{1+3y}-3\lambda _c^\Delta .
\eqry
Note that there are four unknown parameters for each multiplet. By eliminating
them, i.e., $x,y,\lambda _s^\Delta ,\lambda _c^\Delta ,$ from the above nine 
equalities, we can obtain five relations for baryon masses; e.g.,
\bqry
\Omega ^2\;-\;\Delta ^2 & = & 3 \left( \Xi ^{\ast 2}-\Sigma ^{\ast 2}\right) ,
 \\  
\Omega _{ccc}^2-\Delta ^2 & = & 3 \left( \Xi _{cc}^{\ast 2}-\Sigma _c^{\ast
 2}\right) , \\  
\Omega _{ccc}^2-\Omega ^2 & = & 3 \left( \Omega _{cc}^{\ast 2}-\Omega _c^{\ast
 2}\right) ,
\eqry
\bqry
\left( \Sigma _c^{\ast 2}-\Delta ^2\right) + \left( \Omega _c^{\ast 2}-\Xi ^{
\ast 2}\right)  & = & 2\left( \Xi _c^{\ast 2}-\Sigma ^{\ast 2}\right) , \\ 
\left( \Omega _{cc}^{\ast 2}-\Xi _{cc}^{\ast 2}\right) + \left( \Sigma ^{\ast 
2}-\Delta ^2\right)  & = & 2\left( \Xi _c^{\ast 2}-\Sigma _c^{\ast 2}\right) . 
\eqry
However, just four of them are linearly independent, because of an invariance 
of the nine equalities under simultaneous permutation $(x\leftrightarrow y,\;
\lambda _s\leftrightarrow \lambda _c).$

Here only Eq. (25) can be tested, since Eqs. (26)-(29) contain the baryon 
masses not measured so far. For Eq. (25), one obtains (on GeV$^2)$ $1.280\pm 
0.005$ vs. $1.300\pm 0.030,$ taking the electromagnetic mass splittings as a
measure of the uncertainty (since electromagnetic corrections are not included
in our analysis). 

The analysis may be easily repeated for the $J^P=\frac{1}{2}^{+}$ baryons, 
leading to the following two independent mass relations,
\bqry
\left( \Sigma _c^{'2}-N^2\right) + \left( \Omega _c^2\;-\;\Xi ^2\right)  & = &
2\left( \tilde{\Xi }_c^2-\Sigma ^{'2}\right) , \\ 
\left( \Omega _{cc}^2-\Xi _{cc}^2\right) + \left( \Sigma ^{'2}-N^2\right) 
 & = & 2\left( \tilde{\Xi }_c^2-\Sigma _c^{'2}\right) , 
\eqry
where
\bqry
\Sigma ^{'2} & \equiv & a\Lambda ^2+(1-a)\Sigma ^2, \\
\Sigma _c^{'2} & \equiv & b\Lambda _c^2\;+(1-b)\Sigma _c^2, \\
\tilde{\Xi }_c^2\;\! & \equiv & c\;\!\Xi _c^2+(1-c)\;\Xi _c^{'2}
\eqry
are introduced to distinguish between the states having the same flavor content
and $J^P$ quantum numbers, and $a,b,c$ are not known {\it a priori.} In order
to establish the values of $a,b$ and $c,$ we use the following relation for the
intercepts of the $\frac{1}{2}^{+}$ baryon trajectories in the non-charmed 
sector \cite{Kob},
\beq
2\left[ a_N(0)+a_\Xi (0)\right] = 3a_\Lambda (0)+a_\Sigma (0),
\eeq
which has been subsequently generalized to the charmed sector by replacing the
$s$-quark by the $c$-quark, as follows \cite{KT}:   
\beq
2\left[ a_N(0)+a_{\Xi _{cc}}(0)\right] = 3a_{\Lambda _c}(0)+a_{\Sigma _c}(0).
\eeq
It then follows from the corresponding relations based on (1),(2) that,
respectively,
\bqry
\alpha ^{'}_NN^2+\alpha ^{'}_\Xi \Xi ^2 & = & 2\alpha ^{'}_{\Sigma ^{'}}\left(
\frac{3}{4}\Lambda ^2+\frac{1}{4}\Sigma ^2\right),\;\;\;\alpha ^{'}_{\Sigma ^{
'}}\equiv \alpha ^{'}_\Lambda =\alpha ^{'}_\Sigma,  \\
\alpha ^{'}_NN^2+\alpha ^{'}_{\Xi _{cc}}\Xi _{cc}^2 & = & 2\alpha ^{'}_{\Sigma
^{'}_c}\left( \frac{3}{4}\Lambda _c^2+\frac{1}{4}\Sigma _c^2\right),\;\;\;
\alpha ^{'}_{\Sigma _c^{'}}\equiv \alpha ^{'}_{\Lambda _c}=\alpha ^{'}_{\Sigma 
_c},
\eqry
and therefore
\bqry
\Sigma ^{'2} & = & \frac{3}{4}\Lambda ^2+\frac{1}{4}\Sigma ^2, \\
\Sigma ^{'2}_c & = & \frac{3}{4}\Lambda _c^2+\frac{1}{4}\Sigma _c^2,
\eqry
i.e., in the relations (32),(33) $a=b=\frac{3}{4}.$ It is also seen that the
only parameter which is responsible for different weighting of the states
having the same flavor content and $J^P$ quantum numbers is the isospin of the
state. Thus, since both $\Xi _c$ and $\Xi _c^{'}$ have equal isospin $(I=\frac{
1}{2}),$ they should enter a mass relation with equal weights, i.e., in Eq.
(34) $c=1/2,$ and
\beq
\tilde{\Xi }_c^2=\frac{\Xi _c^2+\Xi _c^{'2}}{2}.
\eeq

Equations (25)-(31), with (39)-(41), are new quadratic baryon mass relations.
In the following, we shall make predictions for the baryon masses not measured
so far using these relations.

For the $\frac{1}{2}^{+}$ baryons, in the approximation of equality of the 
slopes in the light quark sector, $\alpha ^{'}_N\cong \alpha {'}_{\Sigma ^{'}}
\cong \alpha ^{'}_\Xi $ (i.e., $\gamma _s^N\alpha ^{'}_N<<1$ in Eqs. (7),(8)),
it follows from (37) that
\beq
2\left( N^2+\Xi ^2\right) \cong 3\Lambda ^2+\Sigma ^2,
\eeq
which is a relation obtained by Oneda and Terasaki in the algebraic approach 
to hadronic physics \cite{OT} which holds with an accuracy of $\sim 1.5$\%:
(in GeV) $5.235\pm 0. 015$ vs. $5.160\pm 0.010.$ Similar approximation for the
$\frac{3}{2}^{+}$ baryons leads, through (2), to relations
\beq
\Omega ^2-\Xi ^{\ast 2}\cong \Xi ^{\ast 2}-\Sigma ^{\ast 2}\cong \Sigma ^{\ast 
2}-\Delta ^2,  
\eeq
which have long been discussed in the literature \cite{DB,OT,JU,BH} and hold 
with a high accuracy, as well as (42). 

The mass of the $\Xi _c^{'}$ can now be obtained from Eqs. (30),(39)-(41).
Using the measured masses of the states entering these relations, one finds
\beq
\Xi _c^{'}=2569\pm 6\;{\rm MeV.}
\eeq
The mass of the $\Omega _c^\ast $ is obtained from (28):
\beq
\Omega _c^\ast =2767\pm 7\;{\rm MeV.}
\eeq
One sees that the value for the $\Xi _c^{'}$ mass (44) lies within the interval
provided by experiment \cite{WA89}. Both (44) and (45) are consistent with the 
values 2580 and 2770 MeV, respectively, predicted by almost all very recent
calculations $[2,\;4,\;5,\;7,\;11].$ 

Now, we have two (independent) relations for the $\frac{3}{2}^{+}$ baryons, 
Eqs. (26) or (27), and (29), to make predictions for the three unknown masses 
of the $\Xi _{cc}^\ast ,$ $\Omega _{cc}^\ast ,$ and $\Omega _{ccc}.$ Similarly,
we have one relation for the $\frac{1}{2}^{+}$ baryons, Eq. (31), to make 
predictions for the two unknown masses of the $\Xi _{cc}$ and $\Omega _{cc}.$ 
In order to obtain two additional relations (for each of the two multiplets), 
we shall use the approximation of equality of the slopes in the light quark 
sector referred to above. Indeed, we have fitted the three, vector meson, octet
baryon, and decuplet baryon mass spectra simultaneously, by using a common 
value of $x$ in Eq. (24) and similar relations for vector mesons and octet 
baryons for all three multiplets. Our results are shown in Table I (the
calculation is completed when $\lambda _s^\Delta $ becomes zero first of the
three $\lambda$'s). It is seen that the best simultaneous fit corresponds to 
$x=0.05\pm 0.01<<1,$ and therefore the approximation of equality of the slopes
in the light quark sector is completely justified.

For the $\frac{1}{2}^{+}$ baryons, it then follows from (7)-(10) (with $\gamma
_s^N\alpha ^{'}_N<<1)$ that
\beq
\alpha ^{'}_\Xi \cong \alpha ^{'}_N,\;\;\;\alpha ^{'}_{\Sigma _c^{'}}\cong
\alpha ^{'}_{\tilde{\Xi }_c}.
\eeq
We now apply the procedure developed for mesons in \cite{BGH1} to baryons, 
using the following relations based on (2) and (46),
\bqryn
\alpha ^{'}_NN^2\;+\;\alpha ^{'}_{\Xi _{cc}}\Xi _{cc}^2 & = & 2\alpha ^{'}_{
\Sigma _c^{'}}\Sigma _c^{'2}, \\
\alpha ^{'}_N\Xi ^2\;+\;\alpha ^{'}_{\Xi _{cc}}\Xi _{cc}^2 & = & 2\alpha ^{'}_{
\Sigma _c^{'}}\tilde{\Xi }_c^2, \\
\frac{1}{\alpha ^{'}_N}\;+\;\frac{1}{\alpha ^{'}_{\Xi _{cc}}} & = & \frac{2}{
\alpha ^{'}_{\Sigma _c^{'}}},
\eqryn
and obtain a sixth power relation for the $\frac{1}{2}^{+}$ baryon masses:
$$\left( \Xi ^2\Sigma _c^{'2}-N^2\tilde{\Xi }_c^2\right) \left(\Xi ^2-N^2
\right) +\Xi _{cc}^2\left( \tilde{\Xi }_c^2-\Sigma _c^{'2}\right) \left( \Xi 
^2-N^2\right) $$
\beq
=4\left( \Xi ^2\Sigma _c^{'2}-N^2\tilde{\Xi }_c^2\right) \left( \tilde{\Xi }_
c^2-\Sigma _c^{'2}\right) .
\eeq
The same procedure applied for the $\frac{3}{2}^{+}$ baryons leads to a
similar sixth power relation for the $\frac{3}{2}^{+}$ baryon masses:
$$\left( \Xi ^{\ast 2}\Sigma _c^{\ast 2}-\Delta ^2\Xi _c^{\ast 2}\right) \left(
\Xi ^{\ast 2}-\Delta ^2\right) +\Xi _{cc}^{\ast 2}\left( \Xi _c^{\ast 2}-
\Sigma _c^{\ast 2}\right) \left( \Xi ^{\ast 2}-\Delta ^2\right) $$ 
\beq
=4\left( \Xi ^{\ast 2}\Sigma _c^{\ast 2}-\Delta ^2\Xi _c^{\ast 2}\right) \left(
\Xi _c^{\ast 2}-\Sigma _c^{\ast 2}\right) .
\eeq
Equations (47) and (48) yield the following values for the masses of the
$\Xi _{cc}$ and $\Xi _{cc}^\ast :$
\bqry
\Xi _{cc} & = & 3610\pm 3\;{\rm MeV,} \\  
\Xi _{cc}^\ast  & = & 3735\pm 17\;{\rm MeV.}  
\eqry
The values for the masses of the $\Omega _{cc}$ and $\Omega _{cc}^\ast $ can
now be obtained from Eqs. (29) and (31), respectively:
\bqry
\Omega _{cc} & = & 3804\pm 8\;{\rm MeV,} \\  
\Omega _{cc}^\ast  & = & 3850\pm 25\;{\rm MeV.}  
\eqry
The remaining value for the $\Omega _{ccc}$ mass is obtained either from (26)
or (27):
\beq
\Omega _{ccc}=\left[ 
\begin{array}{cc}
4930\pm 45\;{\rm MeV} & {\rm from}\;(26), \\
4928\pm 70\;{\rm MeV} & {\rm from}\;(27).
\end{array}
\right.
\eeq
Both results are consistent, as they should be.  

The effect on the $\frac{1}{2}^{+}$ and $\frac{3}{2}^{+}$ baryon spectra of 
setting $x=0$ in Eqs. (24) and corresponding relations for $\frac{1}{2}^{+}$
baryons is negligible $(\leq $ few MeV), except for the splitting between 
nonstrange and singly strange baryons (see (42),(43)). Even in this case the 
absolute size of this splitting is small, and so the included error is not 
more than 2\%. More significantly, this does not affect the multiply strange 
and charm states by more than 1\%.   

Our results are shown in Table II, together with the predictions of many other 
approaches. One sees that our predictions for the charmed baryon masses done 
in the Regge framework are in good agreement with those of different 
approaches. In particular, the predicted value for the $\Xi _c^{'}$ lies in the
range provided by experiment \cite{WA89}, and is in close proximity to 2580 
MeV, consistent with the very recent predictions 
\cite{Sav,RLP,LRP,ZZ,Fran,Jen}. The predicted value for the $\Omega _c^\ast $ 
mass is in close proximity to 2770 MeV, consistent with almost all very recent 
calculations $[1-5,\;7,\;9,\;11].$ 

As remarked by Kaidalov \cite{Kai}, the relations (2),(4), on which our mass
predictions are based, have such a structure that a variation of $\alpha ^{'}_{
kji}$ by 10-15 \% leads only to about 1\% change in the values of masses 
$m_{kji}.$ Thus, although our calculation of the baryon masses in the double-
and triple-charm sectors is based on the assumption of equality of the slopes
in the light quark sector, we expect our results to be insensitive to any 
further adjustment of the values of these slopes.   

Extension of the present framework to the beauty sector, and predictions for
the masses of the beauty baryons will be the subject of a separate publication.

We note (from Table II) with interest that our results are closest to those
derived using a quark-diquark model \cite{ST}. Agreement between such a model
and linear Regge trajectories is expected from both the QCD area law of the
Wilson loop \cite{Sim} and string approach \cite{Egu}. We plan to investigate
this further in the future.

\newpage
\begin{center}
%{\footnotesize 
\begin{tabular}{|c|c|c|c|c|c|c|c|c|c|c|} \hline
x & $\lambda ^\rho _s$ & $\lambda ^N_s$ & $\lambda ^\Delta_s$ & $K^\ast $ &
$\phi $ & $\Sigma ^{'2}$ & $\Xi $ & $\Sigma ^\ast $ & $\Xi ^\ast $ & $\Omega $ 
 \\ \hline
 0     & 0.219 & 0.422 & 0.420 & 900 & 1015 & 1.304 & 1314 & 1392 & 1536 & 1667
 \\ \hline 
 0.010 & 0.209 & 0.407 & 0.395 & 899 & 1015 & 1.302 & 1315 & 1390 & 1534 & 1669
 \\ \hline
 0.020 & 0.201 & 0.392 & 0.371 & 899 & 1016 & 1.299 & 1316 & 1388 & 1533 & 1670
 \\ \hline
 0.030 & 0.192 & 0.377 & 0.348 & 898 & 1017 & 1.296 & 1317 & 1386 & 1532 & 1671
 \\ \hline
 0.040 & 0.182 & 0.363 & 0.326 & 897 & 1017 & 1.295 & 1318 & 1385 & 1531 & 1672
 \\ \hline
 0.050 & 0.175 & 0.350 & 0.305 & 897 & 1018 & 1.293 & 1319 & 1383 & 1530 & 1673
 \\ \hline
 0.060 & 0.167 & 0.336 & 0.285 & 897 & 1018 & 1.291 & 1319 & 1382 & 1529 & 1673
 \\ \hline
 0.070 & 0.159 & 0.323 & 0.266 & 896 & 1018 & 1.289 & 1320 & 1382 & 1529 & 1674
 \\ \hline
 0.080 & 0.152 & 0.310 & 0.249 & 896 & 1019 & 1.287 & 1320 & 1381 & 1529 & 1675
 \\ \hline
 0.100 & 0.137 & 0.286 & 0.215 & 895 & 1019 & 1.284 & 1321 & 1381 & 1529 & 1677
 \\ \hline
 0.150 & 0.104 & 0.232 & 0.140 & 894 & 1019 & 1.281 & 1323 & 1381 & 1529 & 1676
 \\ \hline
 0.200 & 0.075 & 0.185 & 0.077 & 894 & 1019 & 1.280 & 1324 & 1383 & 1530 & 1673
 \\ \hline
 0.277 & 0.038 & 0.122 &  0    & 896 & 1018 & 1.282 & 1323 & 1392 & 1535 & 1667
 \\ \hline
\end{tabular}
%}
\end{center}

{\bf Table I.} Simultaneous fit to the vector meson, octet baryon, and 

\hspace*{0.64in} decuplet baryon spectra, through the relations
\bqryn
\rho ^2 & = & \frac{K^{\ast 2}}{1+x}-\lambda _s^\rho \;=\;\frac{\phi ^2}{1+2x}-
2\lambda _s^\rho , \\
N^2 & = & \frac{\Sigma ^{'2}}{1+x}-\lambda _s^N \;\!=\;\frac{\Xi ^2}{1+2x}-
2\lambda _s^N , \\
\Delta ^2 & = & \frac{\Sigma ^{\ast 2}}{1+x}-\lambda _s^\Delta \;=\;\frac{\Xi 
^{\ast 2}}{1+2x}-2\lambda _s^\Delta \;=\;\frac{\Omega ^2}{1+3x}-3\lambda _s^
\Delta ,
\eqryn

\hspace*{0.64in} as compared to the measured values:
\bqryn
K^{\ast 0}\!\!\! & = & 896\;{\rm MeV,}\;\;\;\phi \;=\;1019\;{\rm MeV,} \\ 
\Sigma ^{'2}\! & = & 1.290\pm 0.003\;{\rm GeV}^2,\;\;\;\Xi \;=\;1318\pm 3\;{\rm
MeV,} \\
\Sigma ^\ast & = & 1385\pm 2\;{\rm MeV,}\;\;\;\Xi ^\ast \;=\;1533.5\pm 1.5\;
{\rm MeV,}\;\;\;\Omega \;=\;1672.5\;{\rm MeV}.
\eqryn

\hspace*{0.64in} The input parameters are:
$$\rho =769\;{\rm MeV,}\;\;\;N=939\;{\rm MeV,}\;\;\;\Delta =1232\;{\rm MeV}.$$

\hspace*{0.64in} $\lambda $'s are measured in GeV$^2.$ 

\bigskip
\begin{center}
{\footnotesize 
\begin{tabular}{|c|c|c|c|c|c|c|c|} \hline
Reference & $\Xi _c^{'}$ & $\Xi _{cc}$ & $\Omega _{cc}$ & $\Omega _c^\ast $ &
$\Xi _{cc}^\ast $ & $\Omega _{cc}^\ast $ & $\Omega _{ccc}$ \\ \hline
Present work & $2569\pm 6$ & $3610\pm 3$ & $3804\pm 8$ & $2767\pm 7$ & $3735\pm
17$ & $3850\pm 25$ & $4930\pm 45$   \\ \hline 
 [1] &      &      & 3737 & 2760 &      & 3797 & 4787   \\ \hline
 [2] & 2579 &      &      & 2768 &      &      &        \\ \hline
 [3] &      &      &      & 2771 &      &      &        \\ \hline 
 [4] & $2580\pm 20$ & $3660\pm 70$ & $3740\pm 80$ & $2770\pm 30$ & $3740\pm 70$
 & $3820\pm 80$ &      \\ \hline
 [5] & 2582 & 3676 & 3787 & 2775 & 3746 & 3851 &        \\ \hline
 [6] &      & 3660 & 3760 &      & 3810 & 3890 &        \\ \hline
 [7] & $2580\pm 10$ &   &   & $2770\pm 10$ &   &   &     \\ \hline
 [8] & $2583\pm 3$ &     &     &     &     &     &      \\ \hline    
 [9] & 2593 &      &      & 2765 &      &      &        \\ \hline
[11] & $2581\pm 2$ &   &   & $2761\pm 5$ &   &   &      \\ \hline
[17] & 2510 & 3550 & 3730 & 2720 & 3610 & 3770 & 4810   \\ \hline
[18] & 2532 &      &      & 2780 &      &      & 5026   \\ \hline
[19] & 2566 & 3605 & 3730 & 2830 & 3680 & 3800 & 4793   \\ \hline
[20] & 2579 & 3645 & 3824 &      & 3733 &      & 4837   \\ \hline
[21] & 2558 & 3613 & 3703 & 2775 & 3741 & 3835 & 4797   \\ \hline
[22] &      & 3710 &      &      & 3750 &      & 4923   \\ \hline
[23] & 2590 &      &      & 2805 &      &      &        \\ \hline
[25] & 2608 &      &      & 2822 &      &      &        \\ \hline
[26] & 2530 & 3511 & 3664 & 2764 & 3630 & 3764 & 4747   \\ \hline
[27] &      &      &      &      &      &      & 5040   \\ \hline
[28] & 2500 &      &      & 2710 &      &      &        \\ \hline
[29] & 2467 &      &      & 2659 &      &      &        \\ \hline
[30] &     &     &     & $2767\pm 35$ &     &     &     \\ \hline
[32] & $2570^{+6+6}_{-3-6}$ &  &  & $2660^{+5+6}_{-3-7}$ &  &  &    \\ \hline
[33] &   & $3630\pm 50$ & $3720\pm 50$ &   & $3735\pm 50$ & $3840\pm 50$ & 
      \\ \hline
[35] &      & 3742 &      &      & 3811 &      &        \\ \hline
[36] & 2570 & 3610 & 3710 & 2740 & 3680 & 3760 & 4730   \\ \hline
[37] & 2596 & 3752 & 3934 & 2811 & 3793 & 3964 & 5127   \\ \hline
[38] & 2600 & 3725 & 3915 & 2811 & 3783 & 3953 & 5106   \\ \hline
[39] & 2690 & 3700 & 3960 & 2810 & 3768 & 3931 & 5019   \\ \hline
[40] & 2616 & 3837 & 4036 &      &      &      &        \\ \hline
[41] & 2583 &      &      & 2772 &      &      &        \\ \hline    
[42] & 2542 & 3710 & 3852 & 2798 & 3781 & 3923 & 5048   \\ \hline
[43] & 2578 & 3661 & 3785 & 2782 & 3732 & 3856 & 4895   \\ \hline
[44] & 2584 & 3758 & 3861 &      &      &      &        \\ \hline
\end{tabular}
}
\end{center}

{\bf Table II.} Comparison of predictions for the charmed baryon masses

\hspace*{0.71in} not measured so far (in MeV).

\hspace*{0.71in}

\hspace*{0.71in} Potential models: \hspace*{0.8in} [1,17-22]

\hspace*{0.71in} Chiral perturbation theory: \hspace*{0.08in} [2]

\hspace*{0.71in} Relativistic quark model: \hspace*{0.24in} [6]

\hspace*{0.71in} Chiral quark model: \hspace*{0.6in} [9]

\hspace*{0.71in} Heavy quark effective theory: [11,35,36]

\hspace*{0.71in} Nonrelativistic quark models: [23,25]

\hspace*{0.71in} Bag models: \hspace*{1.17in} [26-29]

\hspace*{0.71in} Lattice QCD: \hspace*{1.09in} [30,32]

\hspace*{0.71in} QCD spectral sum rules: \hspace*{0.36in} [33]

\hspace*{0.71in} $SU(4)$ skyrmion model: \hspace*{0.36in} [37]

\hspace*{0.71in} Group theoretical models: \hspace*{0.19in} [38,39]

\hspace*{0.71in} Other models: \hspace*{1.03in} [3-5,7,8,40-44]

\end{document}